\documentclass[preprint,authoryear,12pt]{elsarticle}%

\usepackage{amssymb}
\usepackage{graphicx}
\usepackage{epsfig}
\usepackage{amsthm}
\usepackage{longtable}
\usepackage{lscape}

\journal{New Astronomy Review}

\begin{document}

\begin{frontmatter}



\title{Test of depending frequencies of the variable Stars of OC: NGC 6866}


\author[ku]{Gireesh C. Joshi}
\ead{gchandra.2012@rediffmail.com}
\address[ku]{Department of Physics, Kumaun University, Nainital-263002}
\def\astrobj#1{#1}
\begin{abstract}
The search of secondary pulsations is carried out to understand the possible relations among the known parameters of variables of the cluster, NGC 6866. These pulsations arise due to the various ongoing physical phenomena of the variables. Moreover, pulsations of the variables are identified through the visual inspection of their frequency-amplitude distributions. A total of 18 variables among the 28 known variables are showing the secondary pulsation modes. Furthermore, these pulsation modes do not occur for PV, EA, EB, Elliptical and semi-regular. In addition, the field variables seem to be the red-component-stars (RCS) for the studied cluster. The smoothness of the frequency-amplitude curves, signal to noise ratio and the significant limits are play major role for deciding the real peak or frequency values. We are not rejected those amplitude peak of parabolic patterns, for which, the amplitude is greater than then significant limit of variables. The weight of pulsation frequencies is given to be 0.5 for non full cyclic variation but the amplitude is greater than significant limit. Similarly, our present analysis does not support the HADS characteristics of previous known $HADS$ stars within stellar cluster NGC 6866. We are also proposed to new correlation between the secondary pulsation modes (depending frequencies) and the absolute magnitude of known variables. 
\end{abstract}

\begin{keyword}
(Galaxy): Open star cluster; individual: NGC 6866; variable: pulsation; method-data analysis

\end{keyword}

\end{frontmatter}


\def\astrobj#1{#1}
\section{Introduction}\label{s:intro}
Open star clusters (OSC) are the host of various variable stars such as, $\delta$-Scuti, $\gamma$-Doradus, rotational, elliptical etc., and their detailed study is needed to impose the constraint of stellar pulsation models. Since, variables are change their brightness with the time and also contain the various stages of instability strips of the stellar evolution. Moreover, the pulsation variability is arise due to those certain conditions, which are translated into the various instability strips in the Hertzsprung-Russell (HR) or colour–magnitude (CM) diagram \citep{eye08}. Moreover, these pulsations are not found for those stars, which are laying on a fixed instability strip \citep{bri07}. Although, the study of stellar variability of members of the clusters is more effectively used to constraint the model of stellar evolution processes. These constraints depend on the some basic properties and the evolutionary status of individual stellar members. In this connection, the basic parameters of said variables can be derived from the properties of their associated cluster \citep{mow13}. The simultaneous photometric analysis of the observations of variables of a cluster would be required more precise time-series data, which is obtained by the observations of the stars of cluster under the same weather and equal instrumental conditions \citep{kim01}. In the present work, we are utilized the time series data of \cite{jos12} (available at VIZIER service; hereinafter JOS12) for the detailed frequency-amplitude analysis of the variables. The stellar dynamics and evolution study of the Eclipsing binaries are more attractive in the comparison of a single stars. In addition, their period limit is still an open issue, and may need revolutionary revision for the documentation of very short period systems \citep{liu15}. The detection of the stellar pulsation modes of variable and analysis of their associated frequencies may provide a unique opportunity to understand the internal structure of that variable \citep{mow13}. These prescribed pulsation modes are provide important clues for probing the physical properties of variables, including their masses, luminosities, temperatures and metallicities \citep{sal15}. Furthermore, these pulsating variables cover a broad range of the stellar parameters and associated evolutionary stages \citep{der09}. The present manuscript is described about the pulsation search procedure and results within the known variable stars of JOS12. \\
The importance of search of secondary pulsations of variables is briefly described in the Section~\ref{s:pul}. The identification procedure of these secondary pulsations and smoothness procedure of their associated light curves are prescribed in the Sections~\ref{s:tec} and \ref{s:mov} respectively. The multi-periodic analysis of HADS stars have been carried out in the Section~\ref{s:had}. The specific discussion and analysis of the identified WUMA stars of the open cluster, NGC 6866 have been discussed in the Section \ref{s:wum}. In the Section~\ref{s:con}, we are summarized our important results and also prescribed their important features/uses for constraining the model of stellar evolution processes.
\section{Pulsation search in known variable stars}
\label{s:pul}
\begin{figure*}
\includegraphics[width=14.5cm]{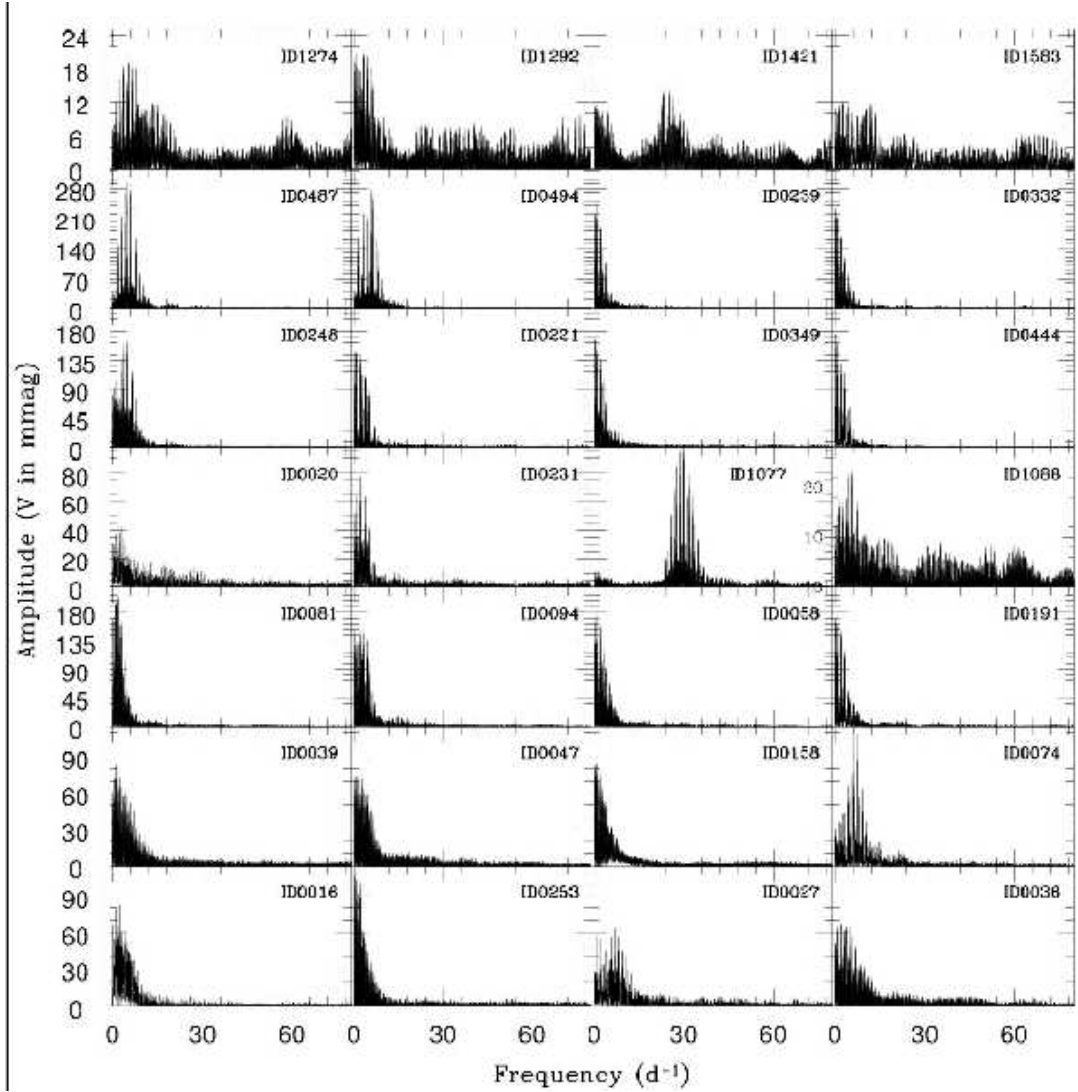}
\caption{The frequency distribution of variable stars in range of 0 to 79.8 $d^{-1}$. The X-axis and Y-axis are represented the frequency ($d^{-1}$) and Amplitude (mmag) in V-band respectively.}
\label{s:fig01}
\end{figure*}

The high variation of reddening may change the measurement accuracy of the distance modulus. The nature of stellar variability does not change through the small variation in stellar magnitude. Since, all science frames of interested target are gathered in the similar environmental conditions, therefore, an analytic view/method is needed to reduce an effect of the positional variation of the target. In addition, the constant environmental conditions are not possible for a 6-8 hours observation of the target in a particular night and also these conditions are varied night to night. Such type effect may reduce through either the simple differential photometry or the secondary standardization method (STM). The transformation equations of Landolt's standard stars are used to generate secondary standard stars in the target field observed on the same night of Landolt's standard field. Secondary standard stars are used to calibrated the instrumental magnitudes on each observing night of target. For this purpose, a linear fit is used between the standard and instrumental magnitudes of these secondary stars assuming that most of the stars are non-variable. Those stars will rejected which deviated by more than $3 \sigma$ deviations. In the differential photometry, the resultant light curves highly scattered due to the adding of the estimation error of the stellar magnitudes. Furthermore, the estimation errors are increasing with the stellar magnitudes and the consideration of comparison star for a known single variable also creates complexity. Since, the STM procedure does not contain these so called facts, therefore, it is used by JOS12 for estimating the stellar magnitudes of members of cluster NGC 6866 from the time-series photometric data. The linear relation of common hundred stars of various frames is a fundamental property of the STM and this procedure may create least scattering due to the many data points. Moreover, we are used archival data (given in VIZIER services) of JOS12 in the present study. JOS12 was observed the cluster for 29 nights between 2008 September 26 and 2011 January 10 (over two observing seasons). According to them, a total of 768 frames in the V-band was accumulated using the 1.04-m Sampurnanand telescope at Manora peak, Nainital and a brief log of their observations are given in Table 1 of JOS12.\\
Since, we are revisited the variable work of JOS12, therefore, the data of light curves of their studied variables is extracted from VIZIER services. These light curves are occurred due to the changing behaviour of the stellar magnitudes. Such magnitude variation of stars arisen due to the various ongoing physical and the stellar evolution phenomenon. Since, each particular phenomena may produce a periodic variation of the stellar magnitudes of studied variable, therefore, the light curves may be contained the various periodic cycles of different amplitude. In the earlier study of these variable, the prominent period (having maximum amplitude) had been determined. As a result, it is further needed to find out other possible periods, which may be produce through the stellar mechanism. In this connection, we are developed a new procedure to find out the secondary pulsation modes on these known variables of the cluster NGC 6866. Since, JOS12 have been reported 28 variable stars in the core region of cluster NGC 6866; therefore, we have been reconstructed the frequency distribution of each variables by using the PERIOD-4.0 program \citep{len05}. This computer program allows us to fit all the frequencies in the given magnitude domain. The amplitude-frequency distribution and primary peak of all prescribed variables are shown in the figure~\ref{s:fig01}.

The amplitude-frequency distribution of these variables is contained the various asymptotically parabolic pattern of the amplitude within the entire frequency domain. These Gaussian profiles of amplitudes are considered to be the possible profiles of the ongoing stellar physical phenomena. The highest peak of amplitude of any asymptotically parabolic pattern is taken as the amplitude of periodic event of the variable and its corresponding frequency is used to estimated the period of said physical phenomena. The major/prominent highest peak of all these asymptotically parabolic pattern may lead as the primary frequency ($f_{0}$) of variables, whereas other amplitudes of remaining asymptotically parabolic pattern may be used to identify the other possible periodic phenomenon. The new prominent frequency of 28 variables and values of their corresponding amplitudes are listed in table~\ref{s:table3}. Moreover, the known literature values (JOS12) of their periods and amplitudes are also prescribed in the same table.

\begin{table*}
\caption{The new frequencies and amplitude values of the variables are listed here with their previous values. Furthermore, they have also classified as RCS and BCS.}
\tiny
\begin{center}
\begin{tabular}{@{}c|c@{}}
\hline\hline
~~~~~~~~~~~~~~~~~~~~~~~~~~~~~~~~~~~JOS12~~~~~~~~~~~~~~~~~~~~~~~~~~~~~~~~~~&~~~~~~~~~~~~~~~~~~Present Study~~~~~~~~~~~~~~~~\\
\hline
 \end{tabular}
 \begin{tabular}{@{}cccccc|ccc@{}}
 \hline
Star & Period & $<V>$ & Amplitude & Membership& Variable & Period & V-Amplitude & Category\\
ID & (d) & (mag) & (V-mag) & status & type & (d) & (mag) & \\
\hline%
0016 &  0.465333 & 11.747 & 0.064 & Likely   & Binary?          & 0.465462 & 0.082 & Blue \\
0020 &  0.380373 & 12.003 & 0.032 & Unlikely & Binary?          & 0.380264 & 0.042 & Blue \\
0027 &  0.143616 & 12.222 & 0.039 & Member   & $\delta$ Scuti   & 0.143638 & 0.064 & Blue \\
0036 &  0.214684 & 12.623 & 0.035 & Likely   & $\delta$ Scuti   & 0.515320 & 0.066 & Blue \\
0039 &  0.836120 & 12.677 & 0.035 & Likely   & $\gamma$ Doradus & 0.835128 & 0.085 & Blue \\
0047 &  0.275710 & 12.995 & 0.041 & Unlikely &   ?              & 0.275606 & 0.079 & Blue \\ 
0058 &  0.911577 & 13.246 & 0.052 & Field    & $\gamma$ Doradus & 0.911635 & 0.182 & Red  \\
0074 &  0.321750 & 13.469 & 0.055 & Likely   & Elliptical       & 0.160855 & 0.117 & Blue \\
0081 &  1.239157 & 13.548 & 0.058 & Likely   & $\gamma$ Doradus & 1.236996 & 0.225 & Blue \\
0094 &  0.740741 & 13.878 & 0.087 & Member   & $\gamma$ Doradus & 0.739284 & 0.168 & Blue \\
0158 &  2.285714 & 14.902 & 0.052 & Member   & EA               & 1.141031 & 0.089 & Blue \\
0191 &  1.090513 & 15.285 & 0.085 & Likely   & Binary?          &20.907380 & 0.173 & Blue \\
0221 &  0.476872 & 15.564 & 0.079 & Likely   & PV               & 0.937075 & 0.153 &  -   \\
0231 &  0.327547 & 15.615 & 0.052 & Member   & PV               & 0.478590 & 0.078 & Blue \\
0239 & 37.037037 & 15.630 & 0.260 & Unlikely & Semi-regular     &37.778617 & 0.283 & Red  \\
0248 &  0.437446 & 15.660 & 0.210 & Likely   & Elliptical       & 0.218657 & 0.166 & Blue \\
0253 &  7.407407 & 15.730 & 0.060 & Field    & Rotational       & 7.554011 & 0.114 & Red  \\
0332 & 11.494253 & 16.190 & 0.140 & Field    & Semi-regular     &11.500862 & 0.237 & Red  \\
0349 & 12.048193 & 16.310 & 0.100 & Unlikely & Semi-regular     &11.870845 & 0.163 & Red  \\
0444 & 16.260162 & 16.820 & 0.310 & Field    & EB               & 8.115565 & 0.177 & Red  \\
0487 &  0.415110 & 17.210 & 0.340 & Member   & W UMa            & 0.207608 & 0.298 & Blue \\
0494 &  0.366704 & 17.260 & 0.440 & Unlikely & W UMa            & 0.183312 & 0.285 &  -   \\
1077 &  0.033559 & 18.580 & 0.340 & Unlikely & HADS             & 0.033559 & 0.100 &  -   \\
1088 &  0.184775 & 18.670 & 0.280 & Field    & HADS             & 0.184864 & 0.024 &  -   \\
1274 &  0.462428 & 19.050 & 0.460 & Unlikely & W UMa            & 0.229491 & 0.020 &  -   \\
1292 &  0.295247 & 18.950 & 0.370 & Unlikely & HADS             & 1.318235 & 0.022 &  -   \\
1421 &  0.041263 & 19.150 & 0.300 & Unlikely & HADS             & 0.041262 & 0.015 &  -   \\
1583 &  0.082055 & 19.260 & 0.320 & Likely   & HADS             & 0.297459 & 0.012 &  -   \\
\hline
\end{tabular}
\end{center}
\label{s:table1}
\end{table*}
\cite{gc16} is used following magnitude-colour relation to separate main sequence (MS) from the stellar distribution of the observed field of cluster,
\begin{equation}
V_{o}=7.66{\times}(B-V)_{o}+7.32,
\end{equation}
where $V_{o}$ and $ (B-V)_{o}$ are the observed magnitude and colour of stars. If, $V_{o}$ value of star satisfy the expression $V_{o} {>} 7.66{\times}(B-V)_{o}+7.32$; then, it is consider to bluer member of studied cluster. The group of bluer member is defined as the blue component stars (BCS) of cluster, which are laying on the left side of main sequence (MS) of the $(B-V)$ vs $V$ colour-magnitude diagram of studied cluster. Other hand, the red component stars (RCS) is group of redder members of stellar cluster and these stars are laying on the right side of the MS of the above prescribed CMD. The redder members of stellar clusters are satisfies the condition of $V_{o} {\leq} 7.66{\times}(B-V)_{o}+7.32$. Our investigate indicates that all members and likely known variables are found to be BCS while all field variables are seem to be RCS. Unlikely members may be either BCS or RCS. JOS12 were assigned various membership status of the variables according to their spatial $P_{sp}$), photometric ($P_{ph}$)  and kinematic probabilities ($P_{pm}$) to assign the stellar membership probabilities of detected stars within observed cluster field. They have chosen these stars as the cluster members either on the basis of photometric probability $P_{ph}=1$ or on the basis of proper motion criteria i.e. kinematic probability ($P_{pm}>0.6$). According to them, Unlikely members are those stars which are satisfied only one criteria and also belong to outside from the core-region ($P_{sp}<0.71$) of the cluster. Those unlikely variables are seem to be BCS, which are selected through the photometric criteria whereas others are RCS. Since, the proper motion distribution of the BCS and RCS does not separate them from each other, therefore, field stars do not identified on the basis of kinematic probabilities.\\
The period of a variable is the reciprocal of the pulsation frequency ($P=1/f$). By keeping in the view of given values of variables by JOS12, this physical definition does not found true for the primary/prominent period of thirteen variables, whom IDs numbers are given as, 0074, 0158, 0191, 0221, 0231, 0248, 0494, 0487, 0494, 1274, 1292 and 1583 respectively. These IDs numbers have been extracted from the list of variables of table~4 of the manuscript of JOS12. Furthermore, variable star of ID 191 has been prescribed as the possible binary star by JOS12, but it is unusual result for this star. Similarly, variable stars of ID 1292 and ID 1583 have identified as the HADS type variables by JOS12, which are not following the above prescribed fact whereas other HADS variable (ID 1077, ID 1088 and ID 1421) are showing good agreement with this fact. The estimated frequency of the contact binary stars is the sum of prominent frequency of each star. Both stars so much near to each other that they are not resolved by the present astronomical telescopes and detected frequency occurs due to their combined physical evolution processes. Let, each star shares equal proportion of mass phenomena of binary system, then the actual frequency of both star will be same. Thus, actual frequency is considered to be the half of the resultant $f_{0}$ of the contact binary stars. On the behalf of present analysis, the known literature values of $f_{0}$ of the PV and EB variables have not justified through the above prescribed statement. These results/facts are also motivated to us for searching the other pulsation frequencies in these known variables.
\section{Technique of identification of new pulsation in variables}
\label{s:tec}
\begin{figure}
\includegraphics[width=14.3cm]{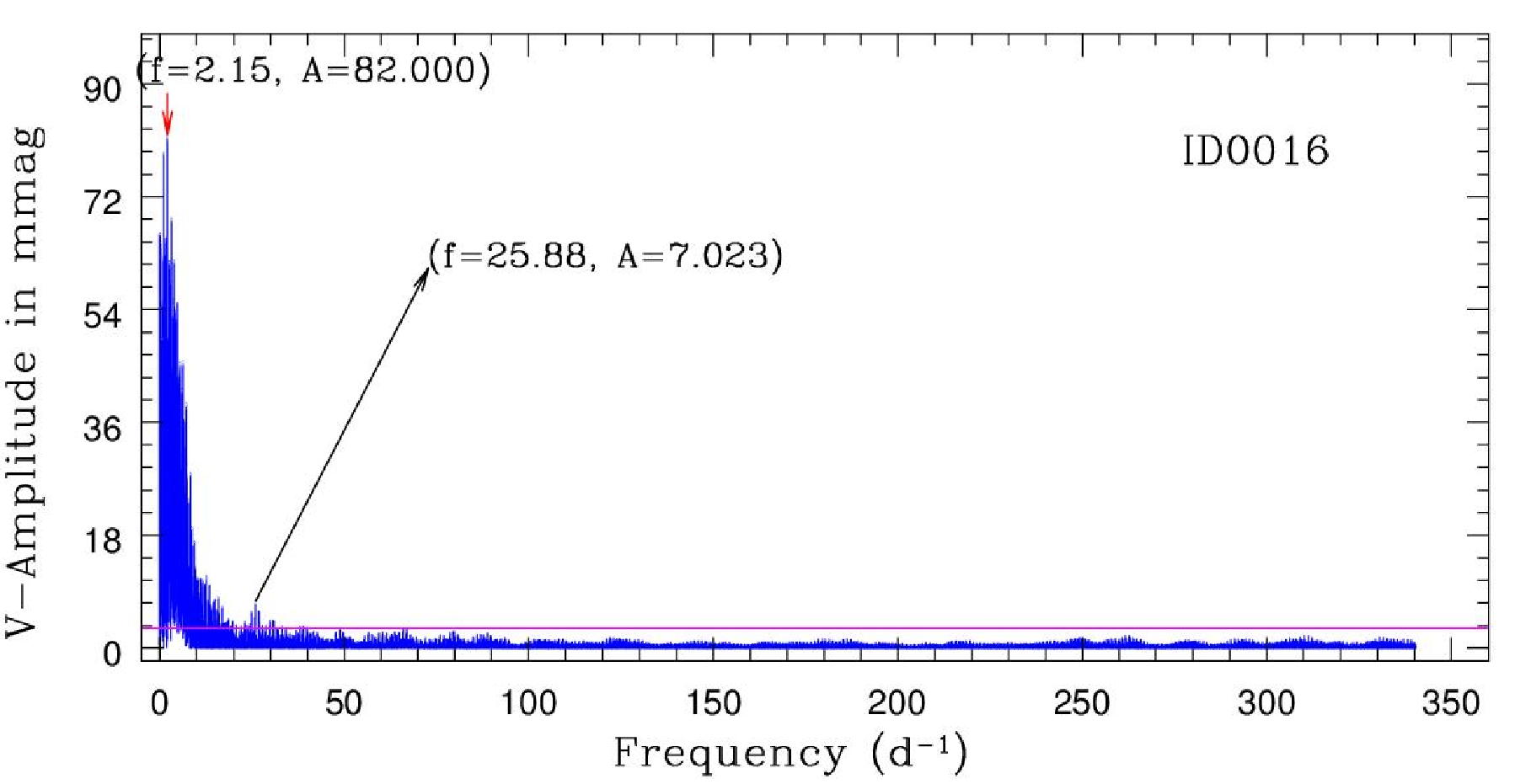}
\includegraphics[width=14.3cm]{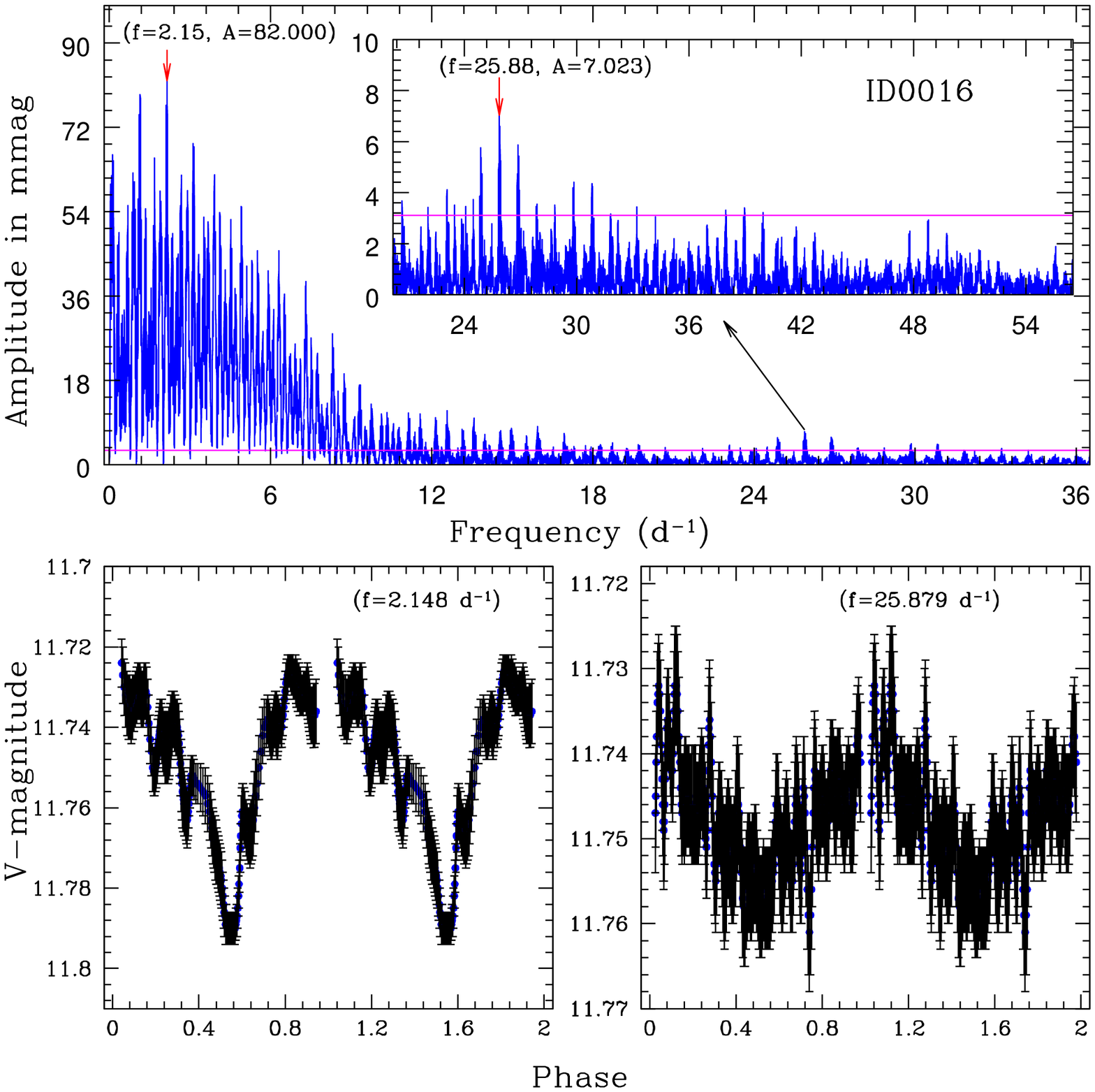} 
\caption{In the upper panel, the amplitude-frequency distribution of variable star having ID 0016 (Range:0-360 $c/d^{-1}$). The X-axis and Y-axis represent the frequency ($d^{-1}$) and Amplitude in V-band (mmag), respectively. In the middle panel, we have shown the frequency distribution of variable star, having ID 0016. The lower panels is depicted the phase diagrams of variable ID 0016 in two frequencies such as,  $2.15~d^{-1}$ and $25.88~d^{-1}$}.
\label{s:fig02}
\end{figure}

The patterns of amplitude-frequency distribution of the variables are appear to be asymptotically parabolic (any-Gaussian, Shannon's function etc.). The high peak of each asymptotic parabola provides the amplitude of that pattern and the position of this highest peak of amplitude provides the value of frequency. The highest amplitude of all such type high peaks considered to be the prominent period and gives the value of ($f_{0}$). It is easily seen that the amplitude of amplitude-frequency diagram of variable is continuously decreasing with the increasing value of frequency until the another asymptotic parabola is not begin. The appeared distribution is also considered to the superimpose of various cyclic variation of ongoing physical phenomenon. For example, we are discussing the nature of amplitude versus frequency pattern of the variable star ID 016, which is a possible Binary. Its amplitude-frequency distribution has been depicted in the figure~\ref{s:fig02}.
In this figure, the amplitude is continuously decreasing within in a frequency domain from $2.15~d^{-1}$ to $20~d^{-1}$ and a asymptotic parabola is appearing. The highest peak of said parabola is found to be $25.88~d^{-1}$. This peak is easily identified on the above prescribed parabola through the visual inspection. In this distribution, we have also found various asymptotic parabola pattern but all these patterns are not prominent with respect to their surrounding. Such type profiles may also occur either due to the instrumental error in the magnitude estimation or due to the stellar amplitude variation by the atmospheric conditions. In addition, the peaks of asymptotic parabolic structures of prescribed diagram are found to be order of $3-4~mmag$ except $7mmag$. Since $7mmag$ is comparatively higher of $3{\sigma}$ limits of remaining peaks of order of $3-4~mmag$, therefore it can not be ignore. On this background, the asymptotically parabolas, having V-amplitude is less than 7 mmag, do not consider for searching the secondary pulsations. Furthermore, they are considered to be the pseudo asymptotically parabolic peaks. On this background, we are found one secondary frequency of variable ID 16. Both frequencies are clearly depicted in the figure~\ref{s:fig02}. The above prescribed figure is also contained the corresponding phase diagrams of variable ID 16. The lower left panel represents the phase diagram through its primary frequency i.e $2.1484~d^{-1}$ while lower right panel represents same for an additional frequency i.e. $25.8787 ~d^{-1}$.\\
At the amplitude spectrum for $f=2.15 d^{-1}$ (upper panel of fig 2), the corresponding amplitude is found to be 0.082 mag. However, the peak to peak complete amplitude from the smoothed phase plots (left lower panel of fig 2)) is computed as 11.792-11.724=0.068 mag i.e. less than 0.082 mag. This result indicates decrement of the amplitude with the data smoothness. The smoothed semi-amplitude of the sinusoid with no harmonics  should be 0.02, so the complete amplitude is to be 2 times of semi-amplitude i.e. 0.04 with no self consistency. Obviously, the light curve at 2.15 $c/d$ shows an eclipse-like narrow minimum, which causes a presence of harmonics of the main frequency, the main at 4.3 $c/d$ and its amplitudes seems larger than that of the ``fast variations". Similarly, the amplitude spectrum shows one more peak at $f=28.88 d^{-1}$. Here, data points strength for $f=2.15 d^{-1}$ is comparatively high rather than data points $f=28.88 d^{-1}$ and characteristics of former has contained more data points rather than later. In this connection, the scattering of phase curve of $f=2.15 d^{-1}$ is found to be due to presence of least characteristic data points of $f=28.88 d^{-1}$. Other hand, the scattering of phase curve of $f=28.88 d^{-1}$ is high due to effect of dominant characteristic of $f=2.15 d^{-1}$. Since, smoothness has been found in the cost of decrement amplitude, therefore, the completed amplitude is found to be $0.026 ~mag  (11.760-11.734)$from the phase plot at $f=25.88 d^{-1}$ (right lower panel of Fig 2). Thus, we have identified a new pulsation frequency of the variable ID 0016 through this procedure.\\
In the present work, we are identified new dependent frequencies of variables. Generally, frequencies of variables could be identify by keeping the consideration of pre-whitened procedure of data-points of light curves. In pre-whitened process, the data points are turn out from the theoretical sine curve of a known amplitude and known pulsation frequency. This process is effective to compute independent pulsation frequencies of the variables, whereas this process reduces the data points for the dependent pulsation frequencies except than prominent pulsation mode. Since, our present study is dedicated to identify the dependent frequencies and their role in the estimation of parameters of variables; therefore, we are not adopted pre-whitened process due to interfered/mixed data-points of dependent frequencies.
\section{Moving average procedure for constructing the phase diagram}
\label{s:mov} 
The light curves are only useful to identify the variable. The real periodic picture of the stellar variability is understandable through the phase-folded diagram. A phase of variable is defined as the measuring scale in the terms of cyclic variation and depends on the period of cyclic variation. Thus, the phase diagrams represents the picture of full cyclic variation of stellar magnitudes. For more clear picture, the points are overlapping in a single phase of all gathered points of all cycles. The phase diagrams of variable ID 0016 (as shown in lower panels of Figure~\ref{s:fig02}) have been constructed through the moving average procedure. In this procedure, the phase of each data point of light curve is derived through the following relation,
\begin{equation}
Phase=(T-T_{0}){\times}f,
\end{equation}
where $T$, $T_{0}$ and $f$ are the JD time of observation epoch, initial JD time i.e. $JD_{0}$ (The JD time of first observed science frame of V-band of NGC 6866 by JOS12 on the date 26 September 2008 i.e. 4736.088461) and corresponding frequency of the variable. The integer part of these computed phase have been removed and remaining decimal part are arranged in the increasing order. As a result, the cyclic variation are overlap to each other and known as the phase-folded diagram. Since, this resultant diagram contained scattered points, therefore, we are needed a procedure to reduce this scattering. Here, we are used moving average algorithm to overcome the effect of scattering. we have been computed the average magnitude of the first 5 data points (1-5) from the prescribed ascended data points. After noting this average value of said points, we are removed first point from the set. Now, we are again computed the average of first five point (2-6 of actual set) and it is noted in the below of previous computed average value. Similarly, this procedure is repeated for upcoming the following sets of data points: 3-7, 4-8 and so on. As a result, we have found a new phase file, in which, a total of 4 data points is less than the number of data points of previous file. This whole procedure have been repeated three times. After the visual inspection of the new resultant phase curve, we have found that the moving average procedure improves the smoothness of phase diagram and also decreases the amplitude of pulsation frequency. The resultant phase diagram of ID 0016 is shown in the Figure~\ref{s:fig02}.
\subsection{Estimation of specific ratio, significant limits and S/N values}
The specific ratio, signal to noise (S/N) values and significant limits of variables are needed to decide the real peak or frequency values. Specific ratios are proposed to co-relate the amplitude of frequency peak ($A(mmag)$) to the mean error of estimation errors of all data points ($Err_{M_{V}}$) of time series photometric data of variables. In the present work, we are computed the specific ratios of variables through the following proposed relation,
\begin{equation}
specefic~ratio=\sqrt{\frac{A(mmag)}{Err_{M_{V}}}},
\end{equation}
The high value of specific ratio represents the smoothness of the phase diagram of the variable. Moreover, this value defines zero for the perfect non-variable stars. Moreover, magnitudes are the logarithm function of flux. Since, the Fourier frequency periodogram (FFP) of light curve of variables shows various asymptotically parabolic patterns of amplitude-frequency variations, therefore, an amplitude label is required to identify the true pulsation frequency. The value of above prescribed label is known as the significant limits. These limits values may be equal or more than $1.5$ times of the noise level. The noise level is the greatest amplitude peak value of identified parabolic frequencies. The background frequencies neither shows the phase curve or does not show the separated and distinguishable asymptotically parabolic patterns of FFP. Since, the $S/N$ value is the ratio of pulsation amplitude to the noise level of FFP, therefore, we are proposed to co-related the $S/N$ value with the significant limit through the following relation,
\begin{equation}
\frac{S}{N}~value \geq \frac{2{\times}A(mmag)}{Significant~limit}.
\end{equation}
We can estimate significant limit for each variable according to its FFP nature . Generally, we are examined the variability nature for those amplitude peak of FFP, which having amplitude more than $1.5$ times of amplitude peak of noise level (As, we shown significant limit about $7mmag$ in Section 3).  The computed specific $S/N$ and significant limits of each variable are prescribed in the table \ref{s:table2}. A deep analysis of Table 4 indicates that the value of $S/N$ ratio of pulsation frequencies is found to be more than 3.2 for variables, having specific ratio greater than 1.4. This prescribed value of $S/N$ is near to the minimum value of Breger criterion. For example, the peak amplitude of background frequencies of parabolic patterns of ID 494 is less than 2.5 mmag for the frequency range from $49.5d^{-1}$ to $340d^{-1}$. Other hand, amplitude peak (4.76 mmag) of asymptotically parabolic pattern of frequencies range $24.5d^{-1}$ to $49.5d^{-1}$ is considered to be the search of possible pulsation frequency. The prescribed amplitude peak ($4.76mmag$) is more than $1.5$ times of $2.5mmag$.\\
The scattering of data-points is played the major role to deciding the significant limits of the amplitude of searched pulsation modes. The scattering is basically depend on the estimation error of photometric magnitudes of the stars; leads the deformation in the shape of light curve of the variables. Such deformation reduces the $S/N$ ratio for the pulsation frequencies. Since, the fainter stars show the higher uncertainty in their computed magnitudes, therefore, the identification of pulsation frequencies of these stars become more complex and less precise. The significant limit may show a chance of identification of pulsation frequencies. Due to the less precision of photometric data, the $S/N$ ratio of secondary pulsation is found to be comparatively low. Since, we are estimated the secondary pulsations of fainter variables on the behalf of significant limit; therefore, the deep and more precise photometric data is needed to confirm these proposed pulsation frequencies, having lower $S/N$.  We are found secondary pulsations for 18 variables of sample of 28 variables. These new identified pulsation frequencies of variables are also listed in the table \ref{s:table4}. 
\begin{table}
\caption{In this table, we have been listed the estimated values of the specific $S/N$ and significant limits of the known variables of the cluster NGC 6866. The errors of pulsation amplitudes is considered to be same mean values of photometric magnitudes as found for the light curve of the variables.}
\tiny
\begin{center}
\begin{tabular}{ccccccc}
\hline
ID   & Error & Amplitude & Specific & Significant & Signal to\\
     & (mag)& (mag) & ratio & limits & noise ratio\\
     &      &    &   &  (mmag) & (S/N)\\\hline
0016 & 0.006 & 0.082 & 3.566 & 3.10 &  52.91\\
0020 & 0.007 & 0.042 & 2.467 & 6.84 &  12.19\\
0027 & 0.007 & 0.064 & 2.969 & 3.48 &  36.57\\
0036 & 0.008 & 0.066 & 2.896 & 5.45 &  24.35\\
0039 & 0.008 & 0.085 & 3.195 & 6.21 &  27.55\\
0047 & 0.009 & 0.079 & 2.984 & 6.89 &  22.95\\
0058 & 0.011 & 0.182 & 4.014 & 5.45 &  66.91\\
0074 & 0.010 & 0.117 & 3.407 & 4.31 &  54.31\\
0081 & 0.011 & 0.225 & 4.428 & 5.14 &  87.68\\
0094 & 0.011 & 0.168 & 3.957 & 4.48 &  74.95\\
0158 & 0.011 & 0.089 & 2.821 & 4.38 &  40.65\\
0191 & 0.013 & 0.173 & 3.649 & 5.02 &  68.81\\
0221 & 0.013 & 0.153 & 3.469 & 5.78 &  52.99\\
0231 & 0.013 & 0.078 & 2.432 & 7.19 &  21.82\\
0239 & 0.013 & 0.283 & 4.689 & 3.73 & 152.01\\
0248 & 0.015 & 0.167 & 3.359 & 2.96 & 112.83\\
0253 & 0.014 & 0.114 & 2.873 & 5.89 &  38.62\\
0332 & 0.016 & 0.237 & 3.804 & 6.16 &  77.08\\
0349 & 0.017 & 0.163 & 3.120 & 5.87 &  55.58\\
0444 & 0.025 & 0.177 & 2.646 & 3.78 &  93.83\\
0487 & 0.026 & 0.298 & 3.399 & 3.57 & 167.09\\
0494 & 0.027 & 0.285 & 3.215 & 2.51 & 227.06\\
1077 & 0.047 & 0.100 & 1.455 & 6.18 &  32.53\\
1088 & 0.074 & 0.024 & 0.570 & 6.65 &   7.23\\
1274 & 0.062 & 0.019 & 0.567 & 5.99 &   6.63\\
1292 & 0.071 & 0.022 & 0.551 & 6.77 &   6.41\\
1421 & 0.060 & 0.015 & 0.495 & 6.53 &   4.53\\
1583 & 0.065 & 0.012 & 0.429 & 6.58 &   3.67\\
\hline
\end{tabular}
\end{center}
\label{s:table2}
\end{table}
\\
A ratio of amplitude signal/noise $\geq$ 4.0 provides a useful criterion for judging the reality of a peak \citep{bre02}. According to $S/N$ values in Table 2, all prominent frequencies values are statistically meaningful and satisfied the Breger criterion. Thus, the computed frequencies are significant. The phase diagrams, having $S/N$ $\geq$ 3.2, indicate smooth and full cyclic variation. Since, fainter stars are lower $S/N$ values compare than Breger criterion, but they are showing full cyclic variation, therefore, their pulsation frequencies are considered by the significant limit rather than Breger criterion. The new identified secondary pulsations are listed in Table 4 and their corresponding $S/N$ is given in the parenthesis of each frequency. However, the secondary frequencies of the fainter stars do not find statistically significant, but they are interesting to study of their dependence in the estimation of parameters of the variables. 
\section{Revisiting towards the Known HADS stars}\label{s:had}
High amplitude Delta Scuti (HADS) variables are those late A and early F type pulsating stars which are changing their absolute magnitude (light) and radial velocity in the periods of one to six hours and also amplitude of these variables is greater than 0.2 mag \citep{bre00}. The values of period and absolute magnitude of these variables are satisfied by the following equations \citep{mcn11},
\begin{equation}
M_{V}=(-2.89{\pm}0.13)log~P-(1.31{\pm}0.10,
\end{equation}
where $M_{V}$ and $P$ are the absolute magnitude and period of the Delta-Scuti stars, which is used by \cite{sal15} for the HADS stars. Similarly, \cite{mcn11} have been also given a relation between $(B-V)_{o}$ and period as given below,
\begin{equation}
(B-V)_{o}=(0.105{\pm}0.004)log~P+(0.336{\pm}0.005,
\end{equation}
where $(B-V)_{o}$ is the colour-index.
In the core field of OSC NGC 6866, a total of five HADS (high amplitude- delta scuti) variables are identified by JOS12. The IDs of these variables are given as 1077, 1088, 1292, 1421 and 1583. All of these given stars have observational $(B-V)_0 > 0.80 mag$, three of them even more than 1 (Table 3). Since, HADS are A-F, main sequence (or slightly more evolved) stars, all of them should have $(B-V)_0 <0.50 mag$; therefore, the classification of HADS stars by JOS12 is very suspicious. Here, we are checking their characteristics by established relationship among the various parameters such as $(B-V)_{o}$, period, absolute magnitude etc. In addition, we are also analyzed the multi-periodicity of these stars as described in the below subsection.
\subsection{ID 1583}
\begin{figure}
\includegraphics[width=10.3cm]{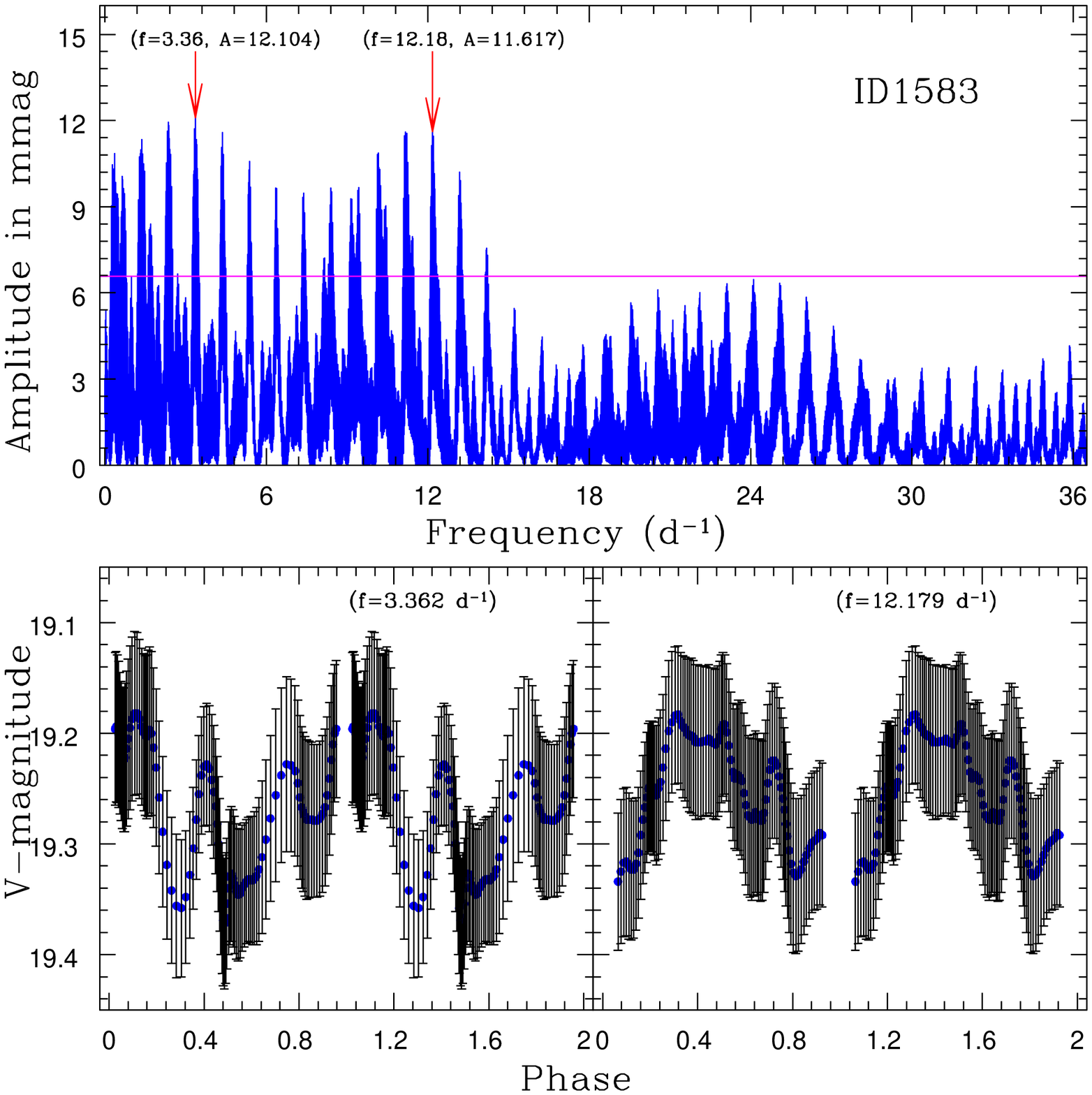}
\includegraphics[width=10.3cm]{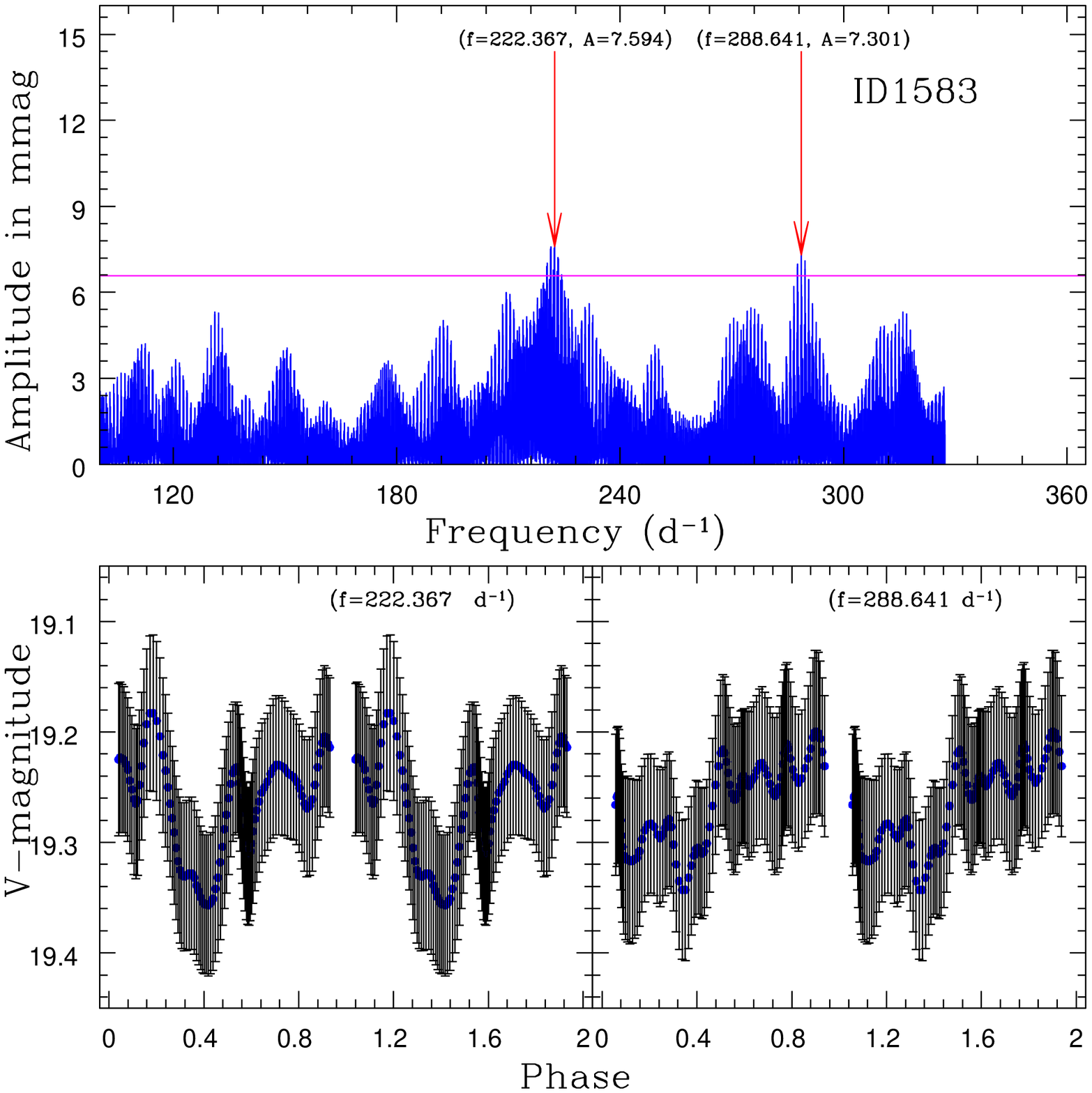}
\caption{The frequency distribution of variable star having ID 1583. The X-axis and Y-axis represent the frequency($d^{-1}$) and Amplitude in V-band respectively. The frequency distribution of variable star (ID 1583) in higher frequency range. The X-axis and Y-axis represent the frequency($d^{-1}$) and Amplitude in V-band respectively. The higher frequency not shown any evidence of full cycle of variability, therefore, it is decline from possible pulsation frequency of variable.}
\label{s:fig15}
\end{figure}

\begin{figure*}
\includegraphics[width=17.3cm]{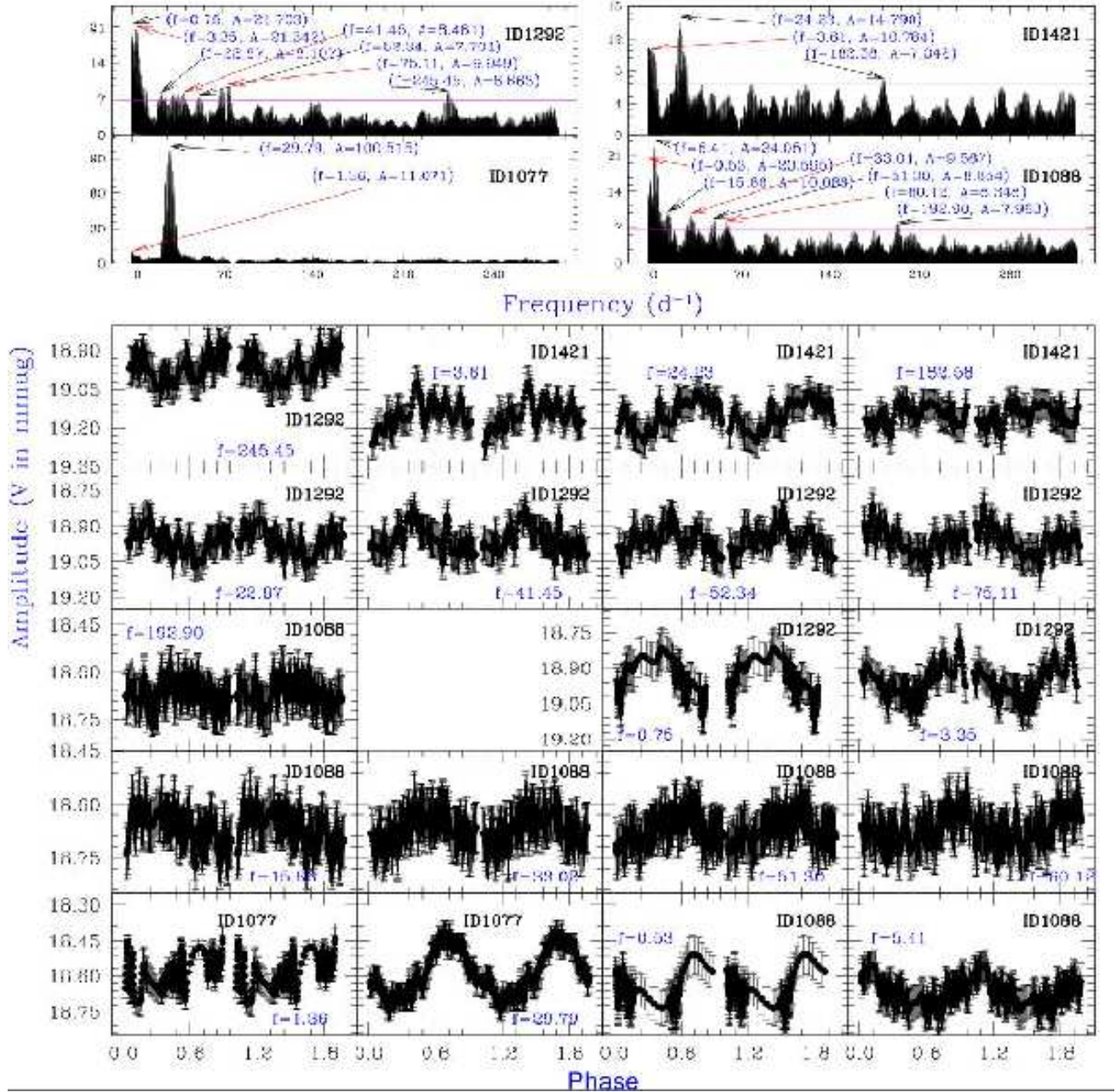}
\caption{The frequency distributions of HADS stars in upper panels. The lower panels represent the corresponding phase diagrams through moving average procedure.}
\label{s:fig16}
\end{figure*}

It is most fainter variable in the field of NGC 6866 and its study is beneficial to understand the nature of cut-off frequency for fainter variables. Its frequency distribution diagram shows four frequencies having Amplitude greater than 7 mmag. There are two frequencies ($3.36~d^{-1}$  and $12.18~d^{-1}$) are found in a single Gaussian distribution and such frequency distribution may classified to be twins frequencies of the variables. The first one is highest amplitude, whereas later is matched with the periodic value of this variable as given by JOS12. These twins frequencies  with their phase diagram are shown in Figure~\ref{s:fig15}.
other two frequencies are found to be  $222.367~days^{-1}$ and $288.641~days^{-1}$, respectively. The position of both frequencies has been shown in the upper panel of Figure~\ref{s:fig16} and corresponding  phase diagrams are depicted in the lower panels of said figure.
It is noticeable fact that the phase diagram of above prescribed first frequency shows a full cyclic variation of the magnitude, while such pattern does not occur for the later frequency. Since, HADS stars are fainter members of the cluster, therefore, their magnitude estimation error is found to be prominent, which may be possible cause of arisen of the later frequency. From all figures of sample stars showing phase diagram, it is apparent that all frequencies have similar amplitudes regardless the particular frequency value. This is physically impossible and makes the identification of higher frequency very suspicious. In addition, frequencies above $100cd_{-1}$ are hardly detectable in the best quality space data or in very fast ground based phtotmetry. This suggests that the detected frequencies may be instrumental nature and may be leads to overestimated results. Other hand, we are not ignored that these frequencies are found for standardized data points of JOS12. In this connection, we gave half weight to this frequency for constraining its relationship with various known parameters.
\subsection{Pulsation Search in other stars}
\begin{figure*}
\includegraphics[width=17.3cm]{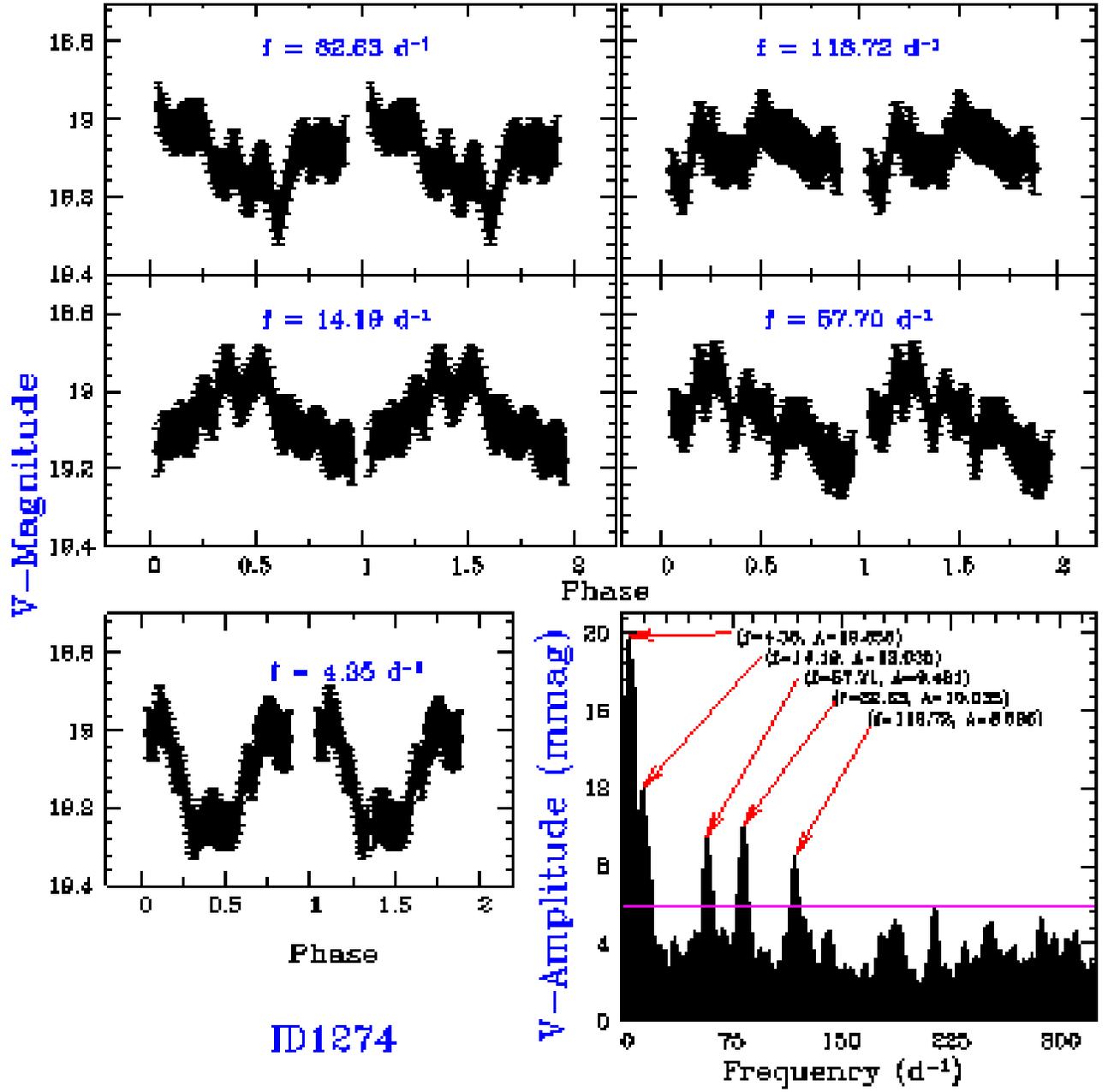}
\caption{The frequency distribution and five pulsation mode of ID 1274.}
\label{s:fig17}
\end{figure*}

We are applied above prescribed approach to identify the pulsation modes in other HADS stars with ID 1274. We are shown these identified frequencies and their corresponding phase diagrams in the Figure~\ref{s:fig17}. Moreover, the phase diagrams of pulsation frequencies of variable ID 1274  are depicted in the Figure~\ref{s:fig18}.
The deep investigation of these diagrams does not show evidence of clear variability for identified frequencies $15.66~d^{-1}$ and  $192.90~d^{-1}$. The amplitude of these frequencies of ID 1088 are 10.088 mmag and 7.963 mmag, respectively. The continuous decremented amplitude does not appear in the both side of these peak values. As a result, such type peaks did not consider for further search of pulsations in other variables. The 182.58 $d^{-1}$ frequency of ID 1421 shows clear variability, which leads our idea about the cut-amplitude of pulsations, which arises through the background effects. In addition, the data points of fainter stars are strongly influenced by the daily aliases and necessarily scattered. The STM has been used for reducing these influence by JOS12. Therefore, the pulsation weight of these frequencies is also considered to be 0.5 in the present case.
\subsection{New modeling for the period and absolute magnitude}
We have found that the fundamental parameters of variables through the cluster studies do not matched with the computed parameters through the well established relation like $M_{V}-P$, $(B-V)_0-P$, Age-Period etc. Since, the emitted photons/light of variable may be result of various ongoing physical phenomenon, therefore, the captured data provides the light curve of dependent pulsation frequencies. On this background, we are added new terms in the well established relation for establishing the similarities   between the parameters through the cluster studies and parameters through the well known relation for independent frequencies. Since, the well known relations like $M_{V}-P$, $(B-V)_0-P$, Age-Period are established for the independent and prominent frequencies of the pulsation variables, therefore, the modified version of these relation is needed for the dependent pulsation frequencies. This modification may be possible through the effective weights and effective number of pulsation frequencies. As a result, we are assumed that every pulsation mode can be effectively contribute to constrain the co-relationship among the various parameters of variable. The actual frequency of contact binaries is taken to be low (half for system of identical stars) compare than the estimated frequency leads to increment in the period. The interfered frequencies (estimated/obtained) is always higher than the prominent frequency. The additional frequencies are affected the prominent frequency in the power form leads the change in the period values. Thus, the modified term $log~P$ of dependent frequencies would be linked to various pulsations as follow,
\begin{equation}
log~P_{ef}=~w_1log~P_p~+~w_2log~P_{s_1}~+....$$ $$+w_nlog~P_{s_n},
\end{equation}
where $w_1,~w_2,...w_n$ are the weight of pulsation modes $P_p,~P_{s_1},.. P_{s_n}$, in which $P_p$ is the period of prominent frequency and $P_{s_1}... P_{s_n}$ are the periodic values of the secondary pulsations. The value of weight is considered to be 1 for the sinusoidal curve of phase, whereas its value is taken to be 0.5 for the half sinusoidal curve of phase.\\
Since, the value of $M_{V}$ of open cluster depends on the value of colour-index, $(B-V)_o$, therefore, we are proposed to add a new term $k~(B-V)_o$ on the \cite{mcn11} relation, in which $k$ is the proportional coefficient of linearity. Since, JOS12 have been categorized some stars as a HADS stars, therefore, we are revisiting these stars by assuming that these stars belongs to same group of variability. In this connection, the value of $k$ is selected to be $3n/2$ for present sample stars through the error and trial method, where $n$ is the number of pulsation modes of prescribed stars. As a result, the new proposed relation becomes as,
\begin{equation}
M_{V}=(-2.89{\pm}0.13)log~P_{ef}+ n_{i}(B-V)_{o}-1.31{\pm}0.10,
\end{equation}
here $n_{i}$ is defined as a sum of weights of detected frequencies i.e.,
\begin{equation}
n_{i}=w_1+w_2+...+w_n.
\end{equation}
Since, above prescribed relation is a statistical relation, therefore, the correlation coefficient of this relation is computed by following relation\footnote{www.real-statistics.com/correlation/multiple-correlation},
\begin{equation}
R_{z,xy}=\sqrt{\frac{r^2_{xz}+r^2_{yz}-2r_{xz}r_{yz}r_{xy}}{1-r^2_{xy}}},
\end{equation}
where $x=log~P_{ef}$, $y=(B-V)_{0}$ and $z=M_{V}$. The correlation between two variable ($r_{xy}$) has been computed through the following relation \citep{ta97},
\begin{equation}
r_{xy}=\frac{\Sigma(x_{i}-\bar{x})(y_{i}-\bar{y})}{\sqrt{\Sigma(x_{i}-\bar{x})^2\Sigma(y_{i}-\bar{y})^2}},
\end{equation}
where $x_{i}$ and $y_{i}$ are the mean values of the variable $x$ and $y$, respectively. The correlation coefficient for absolute magnitude, colour excess and pulsation period is found to be 0.59 through the relation of $R_{z,xy}$. The said value shows strong linear relationship among $M_{V}$, $(B-V)_{0}$ and $log~P$. Furthermore, the corresponding p-value of 5 HADS stars sample is computed to be 0.15 (one-tail test) through the online calculator \citep{so16,co03}.\\
Similarly, the modified relation for computing the colour-excess of $HADS$ as given below,
\begin{equation}
(B-V)_{o}=(0.105{\pm}0.004)log~P_{eff}+n_{i}~(0.336{\pm}0.005.
\end{equation}
The correlation coefficient [$r_{xy}$] of above relation is found to be 0.28,  which does not suggest strong linear dependence between the $(B-V)_{0}$ and $log~P$, but very weak. Thus, the present correlation is statistically incoherent and this analysis cannot give relevant results. However, the p-value of this relation is found to be 0.32 through one-tail test \citep{so16}. A total of 5 sample stars are used to find out the correlations of above prescribed equations. Due to the sample of lack stars (only 5), the statistical test and estimated p-values do not so meaningful to constrain any conclusion for present sample. Present sample stars are also contained very high frequency modes and such modes would definitely be of non-radial nature. The empirical relations for HADS are defined for radial modes and arise from the pulsation equation. These information indicate that HADS categorization by JOS12 is falsified. Thus, we can not make any decision about new proposed empirical relations due to fact that these stars may be completely different stars. Since, their classification seems to be uncertain by statistical algorithm, therefore, we are further investigated their characteristics in the view of values of colour-excess, absolute magnitude and membership probabilities.\\
The resultant model $M_{V}$ and $(B-V)_{o}$ of variables with their observed values are given in the Table~\ref{s:table4}.
\begin{table}
\caption{The model values of absolute magnitude and colour-excess for various HADS stars have been listed here. $I^{st}$ column gives the ID of HADS star. The $n_i$ and $n$ of second column represent the total and effective number of pulsation mode of variables. The abbreviations such as, Mo. and Ob. are the short form of model values and observed values.}
\tiny
\begin{center}
\begin{tabular}{ccccccc}
\hline\hline
Star & $n_{i}$, $n$ & $log~P_{ef}$ & $M_{V}$ & $(B-V)_{o}$ & $M_{V}$ & $(B-V)_{o}$ \\
 ID  &     &  (days) & (Mo.) & (Mo.) & (Ob.) & (Ob.) \\
\hline%
 1088  &  7, 5.5 & -5.4184 & $8.739$ & $1.783$ & $7.92$ & $1.02$\\
 1274  &  5, 4.0 & -5.6254 & $9.267$ & $1.089$ & $8.30$ & $1.42$\\
 1292  &  7, 5.5 & -7.3576 & $8.009$ & $1.579$ & $8.20$ & $1.72$\\
 1421  &  3, 3.0 & -4.2033 & $8.227$ & $0.556$ & $8.40$ & $0.87$\\
 1583  &  4, 3.5 & -5.4184 & $8.913$ & $0.775$ & $8.51$ & $0.99$\\
\hline
\end{tabular}
\end{center}
\label{s:table3}
\end{table}
After deep analysis of Table~\ref{s:table4}, it seems that the model absolute magnitudes of HADS (having kinematic probability less than 0.73 or 0.73) are greater than their observed absolute magnitude, whereas for others HADS members, this fact becomes vice versa. We are found that the model magnitudes seems to be close to each other for last three candidate of Table~\ref{s:table4}, which indicates that they are still cluster members. Other hand, first two candidate show very high deviation between their model and observed values, therefore, they are considered to be field stars by us and these variables are laid within the cluster boundary. The colour excess values through observation and model are also seem to be close for cluster, whereas these described values are far away to each other for field stars. These all $HADS$ stars are fainter member of the cluster, therefore, the photometric procedure/technique of estimation of their apparent stellar magnitude may be showing larger scattering in their light/phase curve. The colour excess values fainter members of cluster seams to be close to each other compare to field stars. Since, these field stars are brighter of the corresponding cluster members, therefore these field stars are either evolved from the cluster region or born different interstellar environment of the cluster region.\\
\subsection{ID 1077: is a SX Phoenicis variable ?}
Due to the spatial position in CMD, it is interesting object for understanding the cluster dynamics. JOS12 are reported that it is a $HADS$ star. $\delta$-Scuti stars (including HADS) are also known as dwarf-Cepheid. Cepheid and Cepheid like variables show the co-relation between age and period. \cite{jos13} are purposed the said co-relation for the Cepheid variables as follow,
\begin{equation}
log(Age)=8.60{\pm}0.07-(0.77{\pm}0.08)~log~P,
\end{equation}
where $P$ is the period of Cepheid. Since, the Classical Cepheids (CCs) shows a double-mode pattern which leads the two peaks in its periodic distribution. Moreover, these two frequency modes are known to be pulsation frequencies of Cepheid or Cepheid like variables. We are also found two frequency modes for ID 1077 as depicted in first two lower panels of Figure~\ref{s:fig16}. The amplitude of its first pulsation mode is low than that of amplitude of second pulsation mode. This peak is symbolic factor of stellar dynamics of the evolution of variable. Since, ID 1077 is a low mass star and it does not satisfied the relation of \citet{jos13}, therefore, it does not a Cepheid. In this connection, the periodic values of ID 1077 does not satisfied the known relation of HADS, therefore, we have been rejected it as a HADS member. However, it shows similar characteristics that of $\delta$-Scuti variables. Since, it shows property of double-mode Cepheid, so we have been examined the \citet{jos13} relation by adding the contributing terms of the dependent secondary pulsation modes. The resultant log(age) does not match with the log(age) of the cluster NGC 6866. The matching of both values of log(age) is needed due to fact that the prescribed ID 1077 is a cluster member. Since, the numerical value of difference of both log(age) is found to be near to reddening value of the cluster, therefore, the colour terms $(B-V)_{o})$ is added in the above prescribed relation. As a result, we are proposed a new relation for ID 1077, which is given as below,
\begin{equation}
log(Age)=8.60{\pm}0.07-(0.77{\pm}0.08)~P_{s}~log~P-(B-V)_{o}.
\end{equation}
where $P_{s}$ is the period of first pulsation mode of the variable ID 1077. The reddened value of colour $(B-V)$ of this variable is $0.66~mag$ (JOS12), which leads the value of $(B-V)_{o}$ as $0.54~mag$. The Amplitude of each pulsation mode of ID 1077 has shown in the corresponding phase diagrams. The above prescribed relation provides the value of log(age) of this variable as $8.89{\pm}0.15$, which shows close agreement with the log(age) of NGC 6866.\\
JOS12 are reported that its periodic is 0.033559 $d$, which is further supplemented by the our present prominent frequency ($29.79d^{-1}$) for ID 1077. This periodic value is satisfied a short period pulsation behaviour of SX Phoenicis variables (i.e. their periodic values varies on time scales of 0.03-0.08 days). The masses of SX Phe variables are in the range 1.0-1.1 $M_{\odot}$ \citep{fi14}. Moreover, such type variables appears more bluer (having a higher temperature) compare to similar luminous stars of the main sequence of studied cluster \cite{sa01}. Since, above prescribed conditions of SX Phoenicis are satisfied for the variable ID 1077, therefore it is a SX Phoenicis type variable of cluster NGC~6866.
\section{WUMA Stars}\label{s:wum}
\begin{figure*}
\includegraphics[width=8cm]{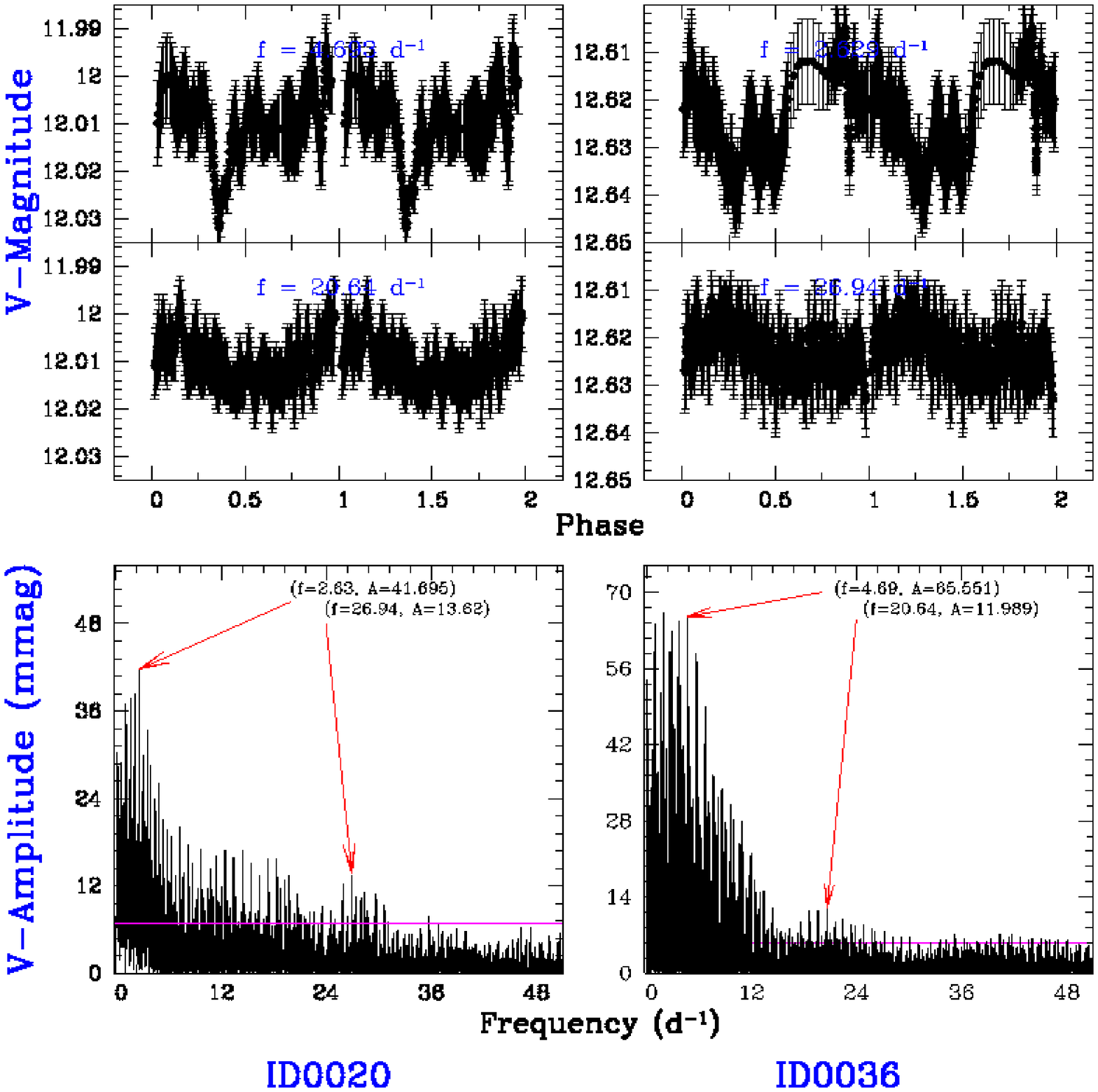}
\includegraphics[width=8cm]{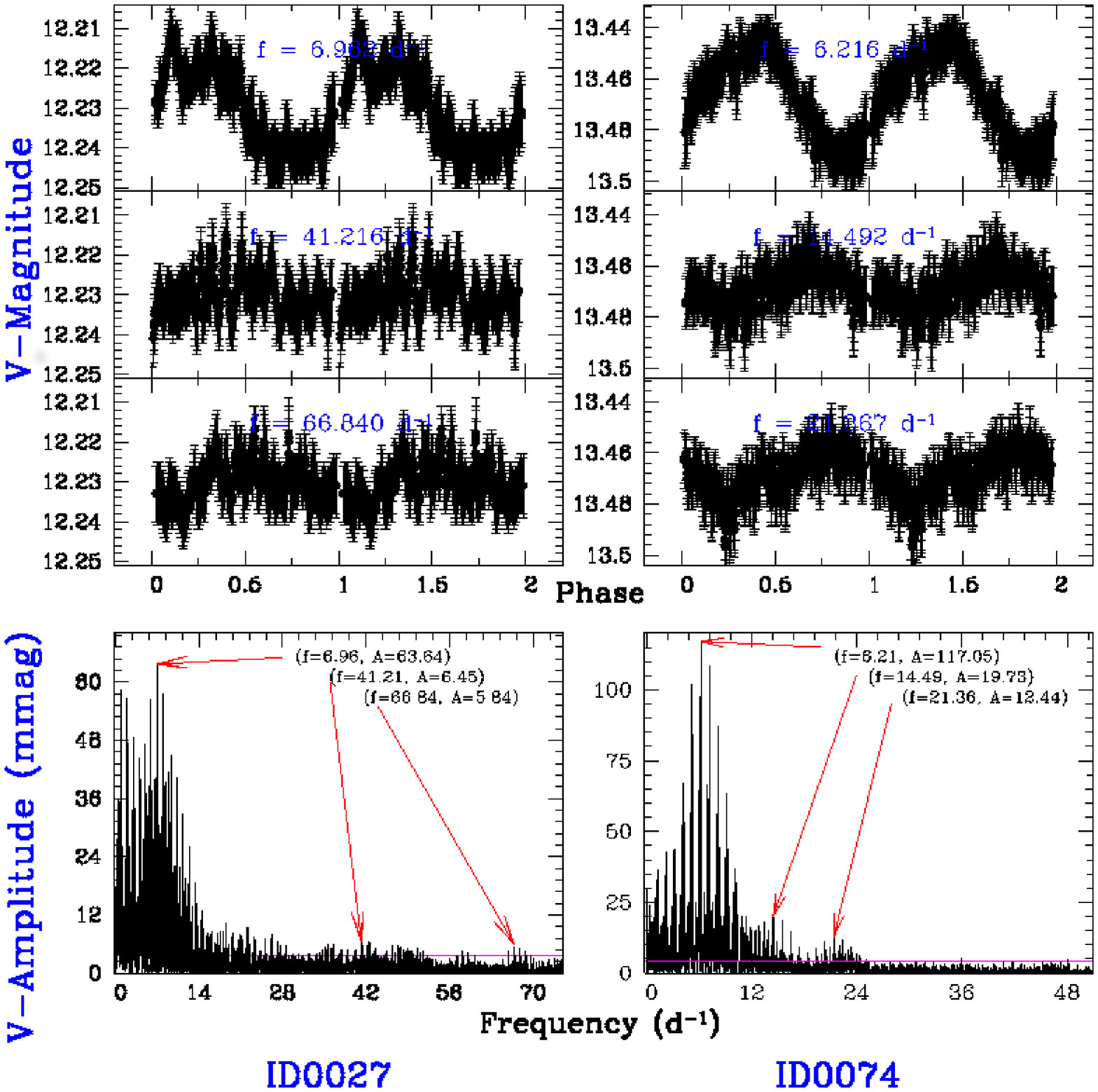}
\includegraphics[width=8cm]{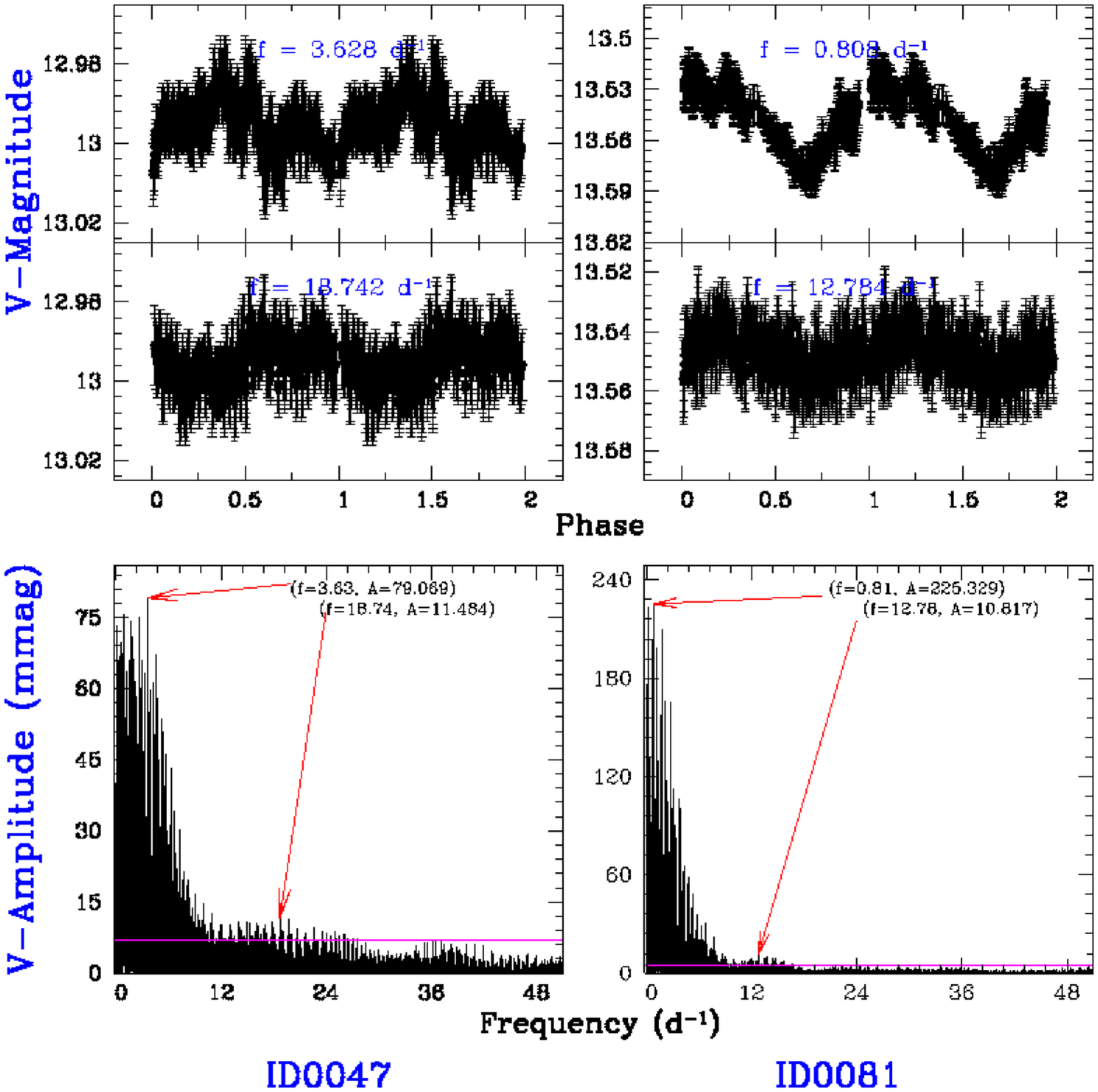}
\includegraphics[width=8cm]{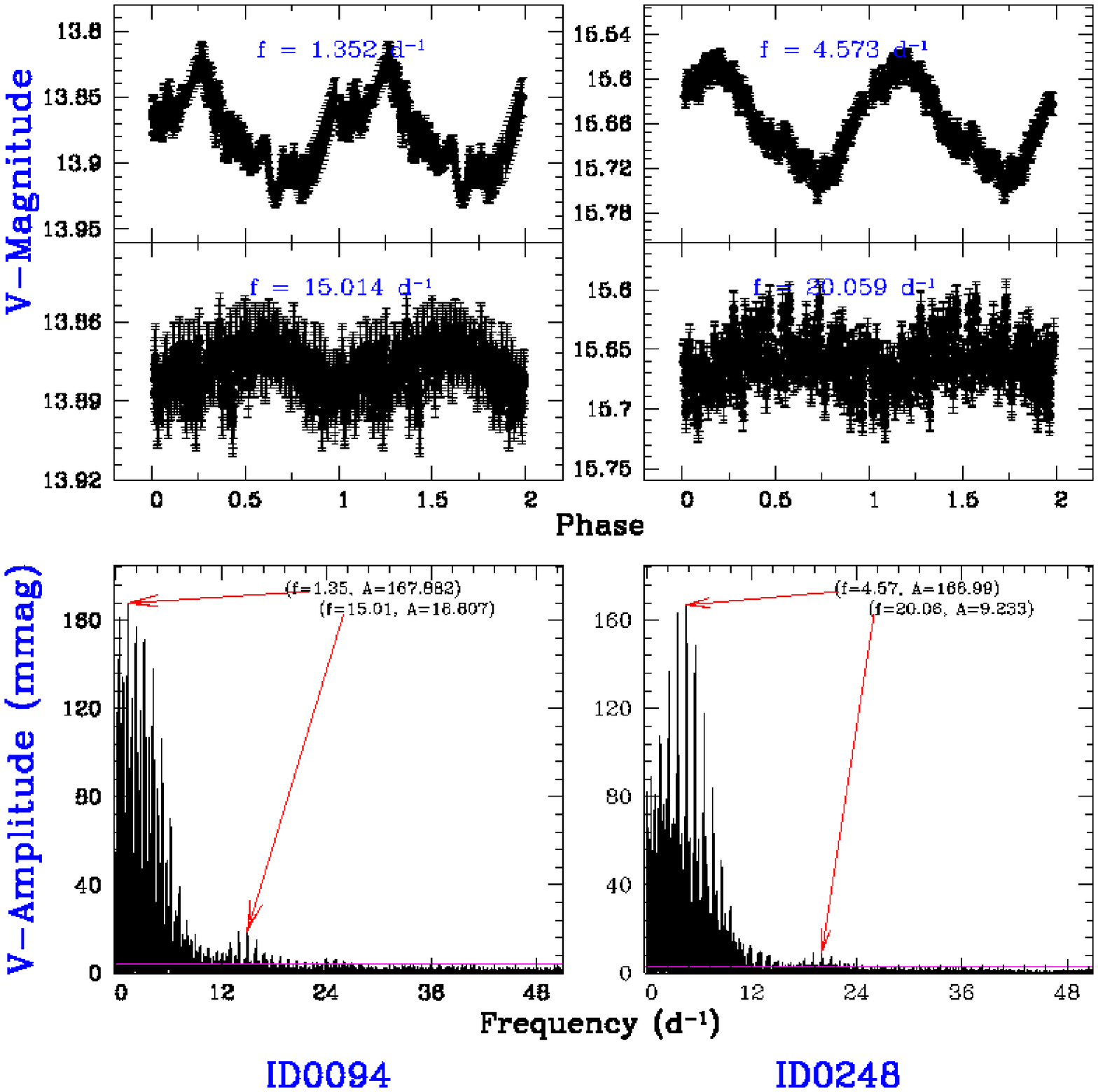}
\caption{The frequency distribution and identified pulsation modes for ID 20, ID 36, ID 27, ID 74, ID 47, ID 81, ID 94 and ID 248.}
\label{s:fig18}
\end{figure*}

\begin{figure}
\includegraphics[width=8cm]{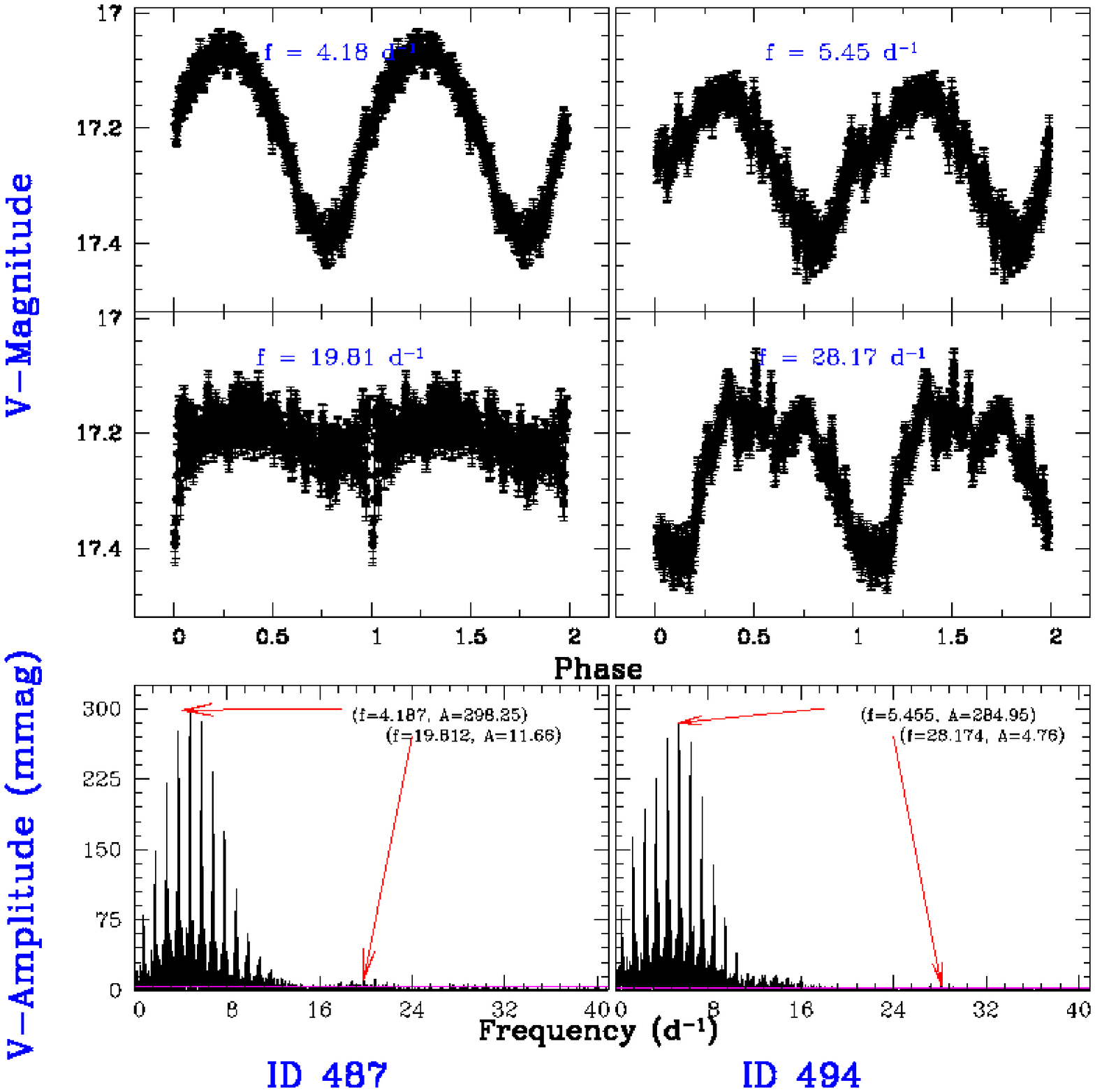}
\caption{The frequency distribution and pulsation mode of ID 487 and ID 494.}
\label{s:fig19}
\end{figure}

A W Ursae Majoris (WUMA) variables are low mass contact binaries, which are a subclass of the eclipsing binary variable stars. The light curves of these variables are contained the continuous brightness variations with the strongly curved maxima and minima of nearly equal depths. JOS12 identified three WUMS stars in the cluster NGC 6866 and their estimated period is listed in the Table 3. For obtaining the true period values of binary systems of ID 487 and ID 494, they had been considered twice of obtained period values. They are argued that EBs are better represented by two sine waves as first author of that paper is accepted in his new paper \citep{josy15}. It is noticeable fact that they did not applied same technique for other possible binary systems ID 16 and ID 20. As a result, they are created a confusion to determine the so called true period of binary systems and does not make a clear scientific reason for obtaining the true period of EBs as a twice of obtained period. The cause of better representation may not be a good scientific reason due to fact that the scattering may be also occurred by the environment effect. Similarly, the position of both variable stars seems to be different in the $(B-V)_{o}$ vs $V$ CMDs of JOS12 and \cite{josy15}. It is highly ethic due to fact that both figures are constructed by same author through same results from same data-sets.\\
We have noticed that first one is depend on the observed stellar magnitude by observation whereas the later one is based on the so called true period of both variables. \cite{josy15} have been computed the true period of ID 487 and ID 494 as $0.415110{\pm}0.000001~d$ and $0.366709{\pm}0.000004~d$, respectively. Furthermore, the absolute magnitude $M_{V}$ of these variables are estimated to be 4.32 mag and 4.86 mag, respectively, through the following \cite{ruc97} relation,
\begin{equation}
M_{V}=-4.44~log(P)+3.02(B-V)_{o}+0.12,
\end{equation}
where $(B-V)_{o}$ and $P$ are the intrinsic colour and orbital period respectively. The shifted values of absolute magnitudes of both variables must be occurred due to the consideration of true period through their assumption.\\
Since, the members of cluster does not changed their position in observed CMD, therefore, we are denying \cite{josy15} procedure for estimating of periods of these variables. Our power spectrum analysis indicates that the prominent period values of ID 487 and ID 494 are found to be $0.207608508~d$ and  $0.183312602~d$, respectively, which leads the absolute magnitude values of these variables as 5.58 mag and 6.14 mag respectively. Other hand, we are obtained these values from the solution of best fitted isochrone as 6.45 mag and 6.51 mag respectively, which are close to our new finding results through \cite{ruc97} relation. However, new periods provide close values of absolute magnitude that of comes through CMD but still less. We are analysis frequency distribution of these variables to know the cause of less absolute  magnitudes through their primary or fundamental period value. For this purpose, we have been carried out a search procedure for identifying the pulsation modes in these variables. On the basis of visual inspection, we are found 5, 2 and 2 pulsation modes in ID 1274, ID 487 and ID 494 respectively. The number of pulsation modes of ID 1274 seems to be more than that of WUMA. Variable ID 1274 studied here due to fact that JOS12 were classified its as a WUMA type variable. Other hand, present analysis indicates that it does not a WUMa. The pulsation modes of ID 1274 are shown in Figure~\ref{s:fig18} and results of its details study is prescribed in the Table \ref{s:table3}.
 The pulsation modes of ID 487 and ID 494 are shown in Figure~\ref{s:fig19}.
 \begin{figure}
\includegraphics[width=8cm]{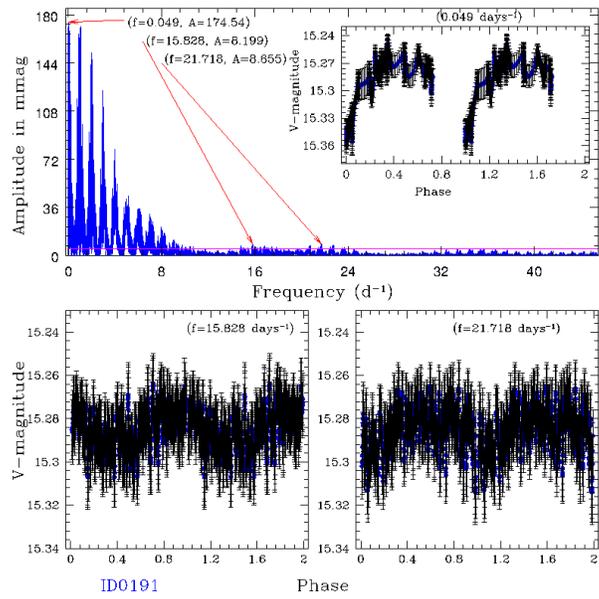}
\caption{The frequency distribution and three pulsation mode in ID 0191.}
\label{s:fig20}
\end{figure}
\begin{table*}
\caption{The identified secondary pulsation peaks for variables in the field of view of NGC 6866.}
\begin{center}
\begin{tabular}{@{}c|c@{}}
\hline\hline
~~~~~~~~~~JOS12~~~~~~~~~~&Pulsation frequencies/ Sub-frequencies (in $d^{-1}$)\\
\hline
 \end{tabular}
 \tiny
 \begin{tabular}{@{}ccc|ccccccc@{}}
 \hline
Star & RA & DEC & $f_1$ & $f_2$ & $f_3$ & $f_4$ & $f_5$ & $f_6$ & $f_7$\\
ID & (J2000) & (J2000) & (S/N) & (S/N) & (S/N) & (S/N) &  (S/N) & (S/N) & (S/N)\\
\hline%
0016 &  20:03:26.12 & 44:10:05.3 & 2.148 & 25.879 & -- & -- & -- & -- & -- \\
     &              &            & (52.91) & (4.52) & -- & -- & -- & -- & -- \\  
0020 &  20:04:25.52 & 44:10:16.2 & 4.693 & 20.689 & -- & -- & -- & -- & -- \\
     &              &            & (12.19) & (3.98) & -- & -- & -- & -- & -- \\
0027 &  20:03:47.13 & 44:09:25.7 & 6.962 & 41.216 & 66.840 & -- & -- &  -- & -- \\
     &              &            & (36.57)& (3.71)& (3.36) & -- & -- &  -- & -- \\
0036 &  20:03:42.47 & 44:10:06.4 & 2.629 & 26.940 & -- & -- & -- & -- & -- \\
     &              &            & (24.35)& (4.40)& -- & -- & -- & -- & -- \\
0047 &  20:04:11.20 & 44:05:33.3 & 3.628 & 18.742 & -- & -- & -- & -- & -- \\
     &              &            & (22.95)& (3.33)& -- & -- & -- & -- & -- \\
0074 &  20:03:34.93 & 44:14:50.1 & 6.216 & 14.492 & 21.367 & -- & -- & -- & -- \\
     &              &            & (54.31)& (9.15)& (5.77) & -- & -- & -- & -- \\
0081 &  20:03:27.93 & 44:09:19.1 & 0.808 & 12.784 & -- & -- & -- & -- & -- \\
     &              &            &(87.68)& (4.21) & -- & -- & -- & -- & --\\
0094 &  20:03:59.34 & 44:10:25.8 & 1.352 & 15.014 & -- & -- & -- & -- & -- \\
     &              &            &(74.95)& (8.39) & -- & -- & -- & -- & -- \\
0191 &  20:03:33.48 & 44:13:53.4 & 0.049 & 15.828 & 21.718 & -- & -- & -- & -- \\
     &              &            & (68.81)& (3.27)& (3.45) & -- & -- & -- & -- \\
0248 &  20:03:38.79 & 44:04:53.0 & 4.573 & 20.059 & -- & -- & -- & -- & -- \\
     &              &            &(112.83)&(6.24) & -- & -- & -- & -- & -- \\
0487 &  20:03:49.82 & 44:11:08.5 & 4.187 & 19.812 & -- & -- & -- & -- & -- \\
     &              &            &(167.09)&(6.53) & -- & -- & -- & -- & -- \\
0494 &  20:04:00.17 & 44:14:03.2 & 5.455 & 28.174 & -- & -- & -- & -- & -- \\
     &              &            &(227.06)& (3.79)& -- & -- & -- & -- & -- \\
1077 &  20:04:13.87 & 44:03:45.8 & 29.797 & 1.366 & -- & -- & -- & -- & -- \\
     &              &            & (32.53) & (3.58)& -- & -- & -- & -- & -- \\
1088 &  20:03:56.20 & 44:12:49.9 & 5.409 & 0.534 & 15.667 & 33.018 & 51.303 & 60.128 & 192.909 \\
     &              &            & (7.23)&(6.19) &(3.03)  & (2.87) & (2.66) & (2.51) & (2.39) \\
1274 &  20:04:26.66 & 44:05:35.9 & 4.358 & 14.198 & 57.708 & 82.633 & 118.726 & -- &  -- \\
     &              &            & (6.63)& (4.02) & (3.17) & (3.35) & (2.87) & &\\
1292 &  20:03:41.23 & 44:12:17.9 & 0.758 & 3.351 & 22.877 & 41.456 & 52.343 &  75.111 & 245.455 \\
     &              &            &(6.41) & (6.30) &(2.39) & (2.50) & (2.27) & (2.94) & (2.56) \\
1421 &  20:03:41.08 & 44:08:47.4 & 3.613 & 24.235 & 182.588 & -- & -- & -- & -- \\
     &              &            & (3.30) & (4.53) & (2.16) & -- & -- & -- & -- \\
1583 &  20:03:58.70 & 44:11:33.5 & 3.362 & 12.179 & 222.367 & 288.641 & -- & -- & -- \\
     &              &            & (3.67) & (3.35)& (2.31) & (2.22) & -- & -- & -- \\
\hline
\end{tabular}
\end{center}
\label{s:table4}
\end{table*}
Since, the power of secondary pulsation mode of ID 487 and ID 494 is very low to compare to its primary pulsation mode but it may be utilized to correction in the \cite{ruc97} relation. The said secondary pulsation mode are interfered with the primary mode but not changed the nature of phase curve due to their low strength. Since, secondary pulsation of these variable is superimposed in its primary mode, therefore, we are adding a new term $2P_S~log(P)$ in the value of $log(P)$ of \cite{ruc97} relation, in which $P_S$ is the period of secondary pulsation mode. The corrected relation is given as below,
\begin{equation}
M_{V}=-4.44~(1+2P_S)~log(P)+3.02(B-V)_{o}+0.12.
\end{equation}
This relation provides the values of absolute magnitude as 5.89 mag and 6.37 mag for ID 487 and ID 494 respectively, which are close to the absolute magnitude as comes through CMD. 
\subsection{Other Results}
JOS12 were mentioned in their manuscript that they are identified seven such variables which having better representing with twice the period given by the Lomb-Scargle periodogram. On the behalf of Section~4.2 of their manuscript, there are also some binary stars, for which, the best fitted period has been considered to be half that of actual ones. In both case, they have argued that the corresponding phase diagram is visually good. The scientific statement is totally absent about the half/twice of the actual period, which is obtained through their data points. It is noticeable in Table 3 that they might be taken the half of period for PV variables, whereas twice for the WUMA, EA, EB and Elliptical type variables. In our present analysis, the variable of ID 191 is also found to be deviated frequency from the their given value and not satisfied their argument like for deviate period of other variables. The frequency amplitude diagram of variable of ID 191 is shown in the Figure 8.

\section{Conclusion}\label{s:con}
The stellar dynamics and the stellar evolution history of the clusters may constrain through the detailed analysis of the identified variables within cluster. The ongoing physical and evolutionary phenomena of the stars are produced various type of variation in their stellar magnitudes. These variation are leading the secondary pulsations of  variables. These stellar variations would be resolved through the temporal analysis of the magnitude-frequency distribution of the time series data of variables. In the present work, we are re-determined the frequency distribution of variables through the available phtotometric data. The peak of maximum amplitude of FFP of variable is considered to be the prominent/principal peak, whereas, other remaining peaks may be secondary pulsation modes. We are not found secondary pulsation modes for EA, EB, PV, rotational and semi-regular type variables. For other variables, the high scattering of data points are appeared in the phase-diagram of each secondary pulse. This scattering may be produced either due to superimpose of wave nature of other pulsation or due to estimation error. Since, the scattering is reduced the sharpness of the characteristics of the variables, therefore, an iterative-moving-average-procedure have been adopted to remove the effect of scattering. The continuous decrements of amplitude of peak has been occurred with the each cycle of this procedure and makes major disadvantage/drawback of this procedure.\\
These secondary pulsation modes are utilized for modifying the relations among various parameters of the HADS and W~UMA type of variable stars. The ID 1274 does not appears WUma, wheres it was claimed as the WUMa by JOS12. The difference in model and observed values of colour [$(B-V)_{o}$] and absolute magnitude for ID 1088 and ID 1274 indicates that both are field stars. We are re-classified ID 1077 as a SX  Phoenicis variable instead of HADS type variable. Since, SX  Phoenicis variable are cousins of the Delta Scuti variables (i.e. dwarf-Cepheid). The variable ID 1077 is also satisfied the age and period relation of Cepheid variables, the result of said relation shows close agreement with the log(age) of the cluster. The number of secondary pulsation modes in the WUMS stars are used to modify the previous known relation, which provides an opportunity to estimate the model absolute magnitude of variables. Furthermore, the present finding of WUMa variables are confirmed by the photometric results of observations. As a result, we are concluded that the investigation of secondary pulsations are open an opportunity to develop their identification techniques and to constrain the models of their arisen. \\
Since, we have been proposed to significant limit for searching the pulsations, therefore, we have been examined the each pulsation of variables, having amplitude peak of asymptotically parabolic pattern is greater than this value. We have been found clear evidence of pulsation for those variables, having specific $S/N$ is greater than 1. For other variables, such type pulsation are highly affected through the background frequencies or noise level. Since, the amplitude of pulsation is less for the HADS stars, therefore their amplitude-frequency distribution is highly influenced through scattering. In addition, peak-amplitude of asymptotically parabolic patterns does not rejected as a pulsation search, having value of specific $S/N < 1$ but having more than significant level. However, such type pulsations do not show clear cycle of pulsations and known as highly influenced pulsations with half weight of unit variation.

\section*{Acknowdege}\label{ss:ack}
This research has made use of the VizieR catalogue access tool, CDS, Strasbourg, France. The original description of the VizieR service was published in A{\&}AS 143, 23. GCJ is also thankful to APcyber Zone (Nanakmatta) for providing computer facilities. GCJ is also thankful to Shree Nilamber Joshi for providing the friendly environment, which becomes a mile stone of my research work.
\bibliographystyle{model1a-num-names}
\bibliography{<your-bib-database>}

\end{document}